\documentclass[usenatbib,usegraphicx,twocolumn]{mn2e}
\usepackage[english]{babel} 
\usepackage{color}

\def\u{\vec{U}}

\def\araa{ARAA}

\def\mnras{MNRAS}

\def\aap{A \& A}

\def\apj{ApJ}

\def\u{{\bf U}} 

\def\V2{V_2}
\def\V2ij{V_{2ij}}
\def\S{{\mathcal S}}
\def\V{\mathcal{V}}
\def\N{{\mathcal N}}

\def\la{\left\langle}

\def\lsim{~\rlap{$<$}{\lower 1.0ex\hbox{$\sim$}}}

\def\gsim{~\rlap{$>$}{\lower 1.0ex\hbox{$\sim$}}}

\usepackage{graphics}
\usepackage{color}
\usepackage{amsmath}
\usepackage{amssymb}
\usepackage{psfrag}
\usepackage{subfig}
\usepackage{graphicx}

\newsavebox{\measurebox}
\begin{document}
\date {} 
\title[Measurement of Galactic synchrotron emission] {The angular power spectrum measurement of the Galactic synchrotron emission in two  fields of the TGSS survey}
\author[S. Choudhuri et al.]{Samir Choudhuri$^{1,2}$\thanks{Email:samir11@phy.iitkgp.ernet.in}, Somnath Bharadwaj$^{2}$, Sk. Saiyad Ali$^{3}$, Nirupam Roy$^{4}$
\newauthor Huib.~T.~Intema$^{5}$ and Abhik Ghosh$^{6,7}$\\
$^{1}$ National Centre For Radio Astrophysics, Post Bag 3, Ganeshkhind, Pune 411 007, India\\
  $^{2}$ Department of Physics,  \& Centre for Theoretical Studies, IIT Kharagpur,  Kharagpur 721 302, India\\
$^{3}$ Department of Physics,Jadavpur University, Kolkata 700032, India\\
$^{4}$ Department of Physics, Indian Institute of Science, Bangalore 560012, India\\
$^{5}$ Leiden Observatory, Leiden University, Niels Bohrweg 2, NL-2333CA, Leiden, The Netherlands\\
$^{6}$ Department of Physics and Astronomy, University of the Western Cape, Robert Sobukwe Road, Bellville 7535, South Africa\\
$^{7}$ SKA SA, The Park, Park Road, Pinelands 7405, South Africa
}
\maketitle

\begin{abstract}
Characterizing the diffuse Galactic synchrotron emission at arcminute 
angular scales is needed to reliably remove foregrounds in
cosmological 21-cm measurements. The study of this emission is also
interesting in its own right. Here, we quantify the fluctuations of
the diffuse Galactic synchrotron emission using visibility data for
two of the fields observed by the TIFR GMRT Sky Survey (TGSS).  We have
used the 2D Tapered Gridded Estimator (TGE) to estimate the angular
power spectrum $(C_{\ell})$ from the visibilities. We find that the
sky signal, after subtracting the point sources, is likely
dominated by the
diffuse Galactic synchrotron radiation across the angular multipole
range $240 \le \ell \lesssim 500$. 
 We present a power law fit,
$C_{\ell}=A\times\big(\frac{1000}{l}\big)^{\beta}$, to the measured
$C_{\ell}$ over this $\ell$ range. We find that $(A,\beta)$ have values 
$(356\pm109~{\rm mK^2},2.8\pm0.3)$ and 
$(54\pm26~{\rm mK^2},2.2\pm0.4)$ in the two fields. 
For the second field, however,   there is indication of a significant
residual point source contribution, and for this field we interpret the measured $C_{\ell}$
as 
an upper limit for the  diffuse Galactic synchrotron emission.
While in both fields the slopes are 
consistent with earlier measurements, the second field appears 
to have an amplitude which is considerably smaller compared to similar  
measurements  in other parts of the sky.

\end{abstract} 

\section{Introduction}
Observations of the redshifted 21-cm signal from the Epoch of
Reionization (EoR) contain a wealth of cosmological and astrophysical
information
\citep{bharadwaj05,furlanetto06,morales10,pritchard12}. The Giant
Metrewave Radio Telescope
(GMRT, \citealt{swarup})
is currently functioning at a frequency band which corresponds to
 the 21-cm
signal from this epoch. Several ongoing and future experiments such as
the Donald C. Backer Precision Array to Probe the Epoch of
Reionization (PAPER,
\citealt{parsons10}), the Low Frequency Array
(LOFAR, \citealt{haarlem}), the
Murchison Wide-field Array
(MWA, \citealt{bowman13}), the
Square Kilometer Array (SKA1
LOW, \citealt{koopmans15})
and the Hydrogen Epoch of Reionization Array
(HERA, \citealt{neben16}) are
aiming to measure the EoR 21-cm signal. The EoR 21-cm signal is
overwhelmed by different foregrounds which are four to five orders of
magnitude stronger than the expected 21-cm signal
\citep{shaver99,ali08,ghosh1,ghosh2}. Accurately modelling and
subtracting the foregrounds from the data are the main challenges for
detecting the EoR 21-cm signal. The diffuse Galactic synchrotron emission (hereafter, DGSE) is
expected to be the most dominant foreground at  $\>10$
arcminute angular scales after point source subtraction at {\rm 10-20 mJy} level
\citep{bernardi09,ghosh12,iacobelli13}.  A precise characterization
and a detailed understanding of the DGSE  is
needed to reliably remove foregrounds in 21-cm experiments. In this
paper, we characterize the DGSE at
arcminute angular scales which are relevant for the cosmological 21-cm studies.

The study of the DGSE  is also
important in its own right. The angular power spectrum ($C_{\ell}$) of
the DGSE quantifies the fluctuations
in the magnetic field and in the electron
density of the turbulent interstellar medium (ISM) of our Galaxy
(e.g. \citealt{Waelkens,Lazarian,iacobelli13}).

There are several observations towards characterizing the DGSE spanning a wide range of frequency.
\citet{haslam82} have measured the all sky diffuse Galactic
synchrotron radiation at ${\rm 408~MHz}$. \citet{reich82} and
\citet{reich88} have presented the Galactic synchrotron maps at a
relatively higher frequency $({\rm 1420~ MHz})$. Using the ${\rm 2.3~GHz}$
Rhodes Survey, \citet{giardino01} have shown that the $C_{\ell}$ of
the diffuse Galactic synchrotron radiation behaves like a power law
$(C_{\ell}\propto\ell^{-\beta})$ where the power law index $\beta=2.43$ in the $\ell$
range $2\le\ell\le100$.  \citet{giardino02} have found that the value of
$\beta$ is $2.37$ for the ${\rm 2.4~GHz}$ Parkes Survey in the $\ell$
range $40\le\ell\le250$. The $C_{\ell}$ measured from the {\it
  Wilkinson Microwave Anisotropy Probe} (WMAP) data show a slightly
lower value of $\beta$ $(C_{\ell}\propto\ell^2)$ for $\ell<200$
\citep{bennett03}. \citet{bernardi09} have analysed ${\rm 150~MHz}$
Westerbork Synthesis Radio Telescope (WSRT) observations to
characterize the statistical properties of the diffuse Galactic
emission and find that
\begin{equation}
C_{\ell}=A\times\big(\frac{1000}{l}\big)^{\beta} {\rm mK^2}
\label{eq:eq1}
\end{equation}
where $A=253~{\rm mK^2}$ and $\beta=2.2$ for
$\ell\le900$. \citet{ghosh12} have used GMRT ${\rm 150~MHz}$
observations to characterize the
foregrounds for 21-cm experiments and find that
 $A=513~{\rm mK^2}$ and $\beta=2.34$
in the $\ell$ range $253\le\ell\le800$. Recently, \citet{iacobelli13}
present the first LOFAR detection of the DGSE  around ${\rm 160~MHz}$.  They reported that the $C_{\ell}$ of
the foreground synchrotron fluctuations is approximately a power law
with a slope $\beta\approx1.8$ up to angular multipoles of ${\rm
  1300}$.

In this paper we study the statistical properties of the DGSE using two fields observed by the TIFR
GMRT Sky Survey
(TGSS{\footnote{http://tgss.ncra.tifr.res.in}}; \citealt{sirothia14}).
We have used the data which was calibrated and processed by
\citet{intema16}. We have applied the
Tapered Gridded Estimator (TGE; \citealt{samir16b}, hereafter Paper I) to the residual
data to measure the $C_{\ell}$ of the background sky signal after
point source subtraction. The TGE suppresses the contribution from the
residual point sources in the outer region of the telescope's field of view (FoV) and
also internally subtracts out the noise bias to give an unbiased
estimate of $C_{\ell}$ \citep{samir16a}. For each field we are able to
identify an angular multipole range where the measured $C_{\ell}$ is likely
dominated by the DGSE, and we present power
law fits for these.

\section{Data Analysis}
\label{analysis}
The TGSS survey
contains 2000 hours of observing time  divided on 5336
individual pointings on an approximate hexagonal grid. The observing
time for each field is about ${\rm 15}$ minutes.  For the purpose of
this paper, we have used only two data sets for two fields located at Galactic
coordinates $(9^{\circ},+10^{\circ}$; {\bf Data1}) and
$(15^{\circ},-11^{\circ}$; {\bf Data2}). We have selected these fields because they are close to the Galactic plane, and also the contributions from the very bright compact sources are much less in these fields. The central frequency of this
survey is ${\rm 147.5~MHz}$ with an instantaneous bandwidth of ${\rm
  16.7~MHz}$ which is divided into $256$ frequency channels. All the TGSS
raw data was analysed with a fully automated pipeline based on the
SPAM package \citep{intema09,intema14}. The operation of the
SPAM package is divided into two parts: (a){\it Pre-processing}
and (b) {\it Main pipeline}. The Pre-processing step calculates good-quality instrumental calibration from the best
available scans on one of the primary calibrators, and transfers these
to the target field. In the Main pipeline the direction independent
and direction dependent calibrations  are calculated for each field, 
and the calibrated visibilities are converted into 
``CLEANed'' deconvolved radio images. The off source rms noise
($\sigma_n$) for the continuum images of these fields are $4.1~{\rm mJy}/{\rm Beam}$ and
$3.1~{\rm mJy}/{\rm Beam}$ for {\bf Data1} and {\bf Data2}
respectively, both values lie close to the median rms. 
noise  of $3.5~{\rm mJy}/{\rm Beam}$ for the whole survey. The angular resolution of these observations is  ~$25^{''} \times 25^{''}$. This
pipeline applies direction-dependent gains to  image and subtract 
point  sources to a $ S_c=5 \, \sigma_n$ flux threshold covering an angular region 
of  radius  $\sim1.5$ times the telescope's FoV ($3.1^{\circ} \times 3.1^{\circ}$), 
and also includes a  few bright sources even further away. The 
subsequent analysis here uses the  residual visibility data after 
subtracting out the discrete sources.

\begin{figure*}
\begin{center}
\psfrag{cl}[b][t][1.5][0]{$C_{\ell} ~~{\rm mK}^2$}
\psfrag{l}[c][c][1.5][0]{$\ell$}
\psfrag{tot}[r][r][1][0]{Total}
\psfrag{res}[r][r][1][0]{Residual}
\includegraphics[width=80mm,angle=0]{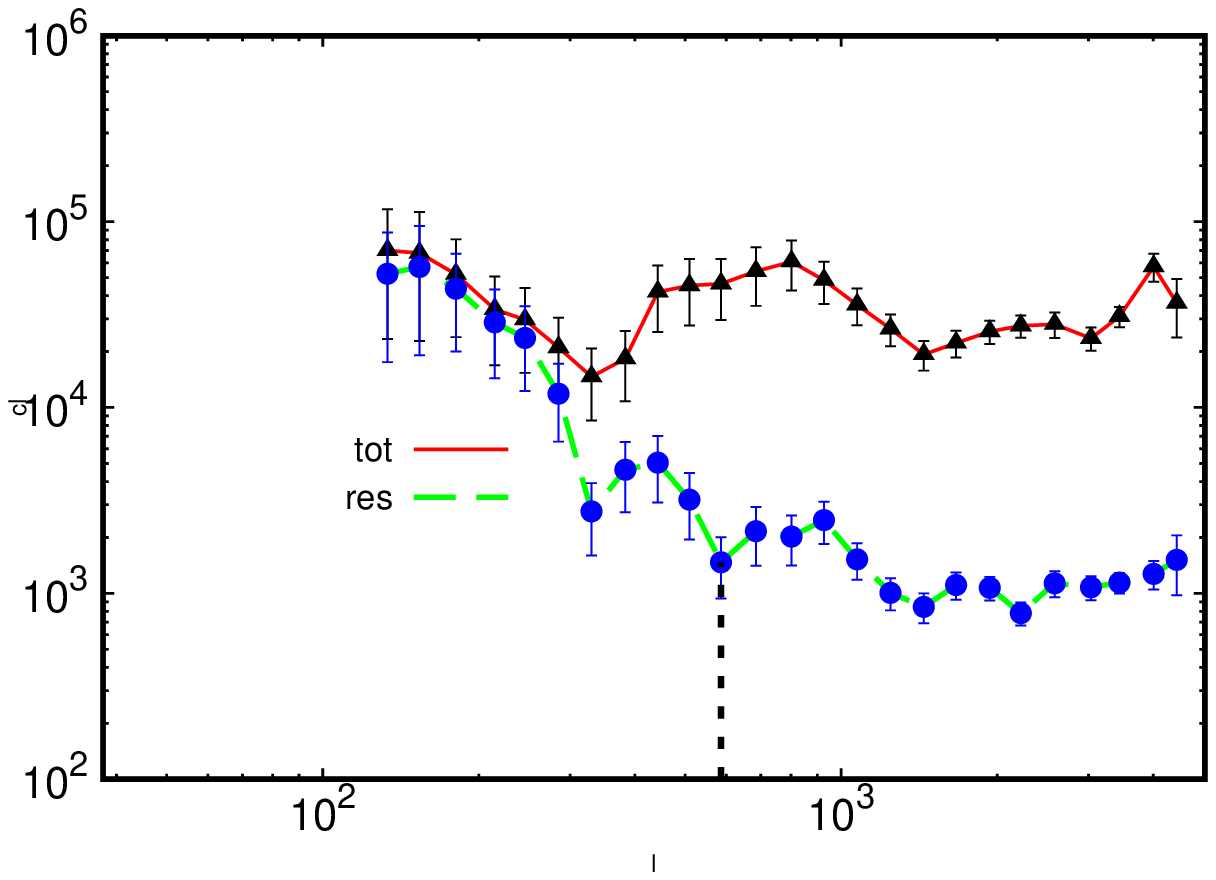}
\includegraphics[width=80mm,angle=0]{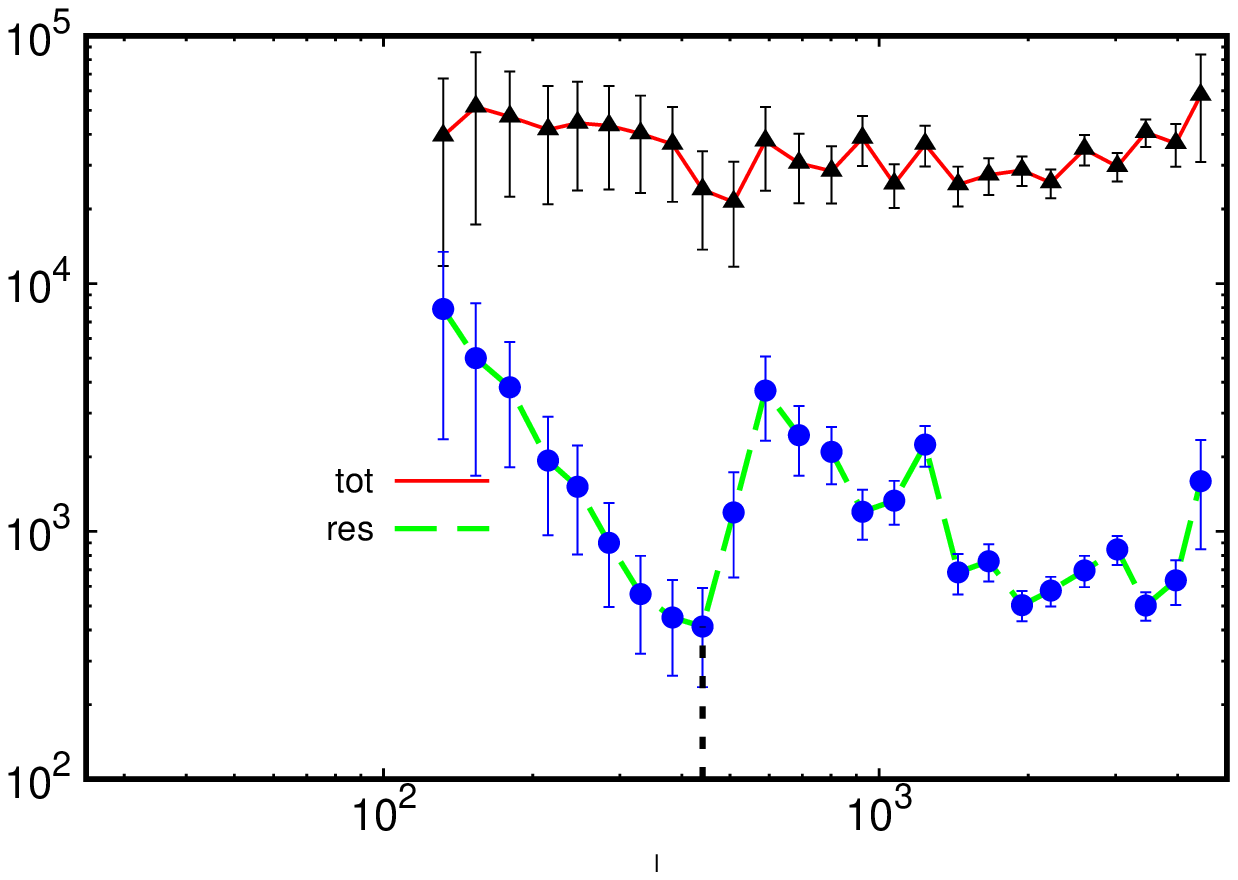}
\caption{Estimated angular power spectra $(C_{\ell})$ with $1-\sigma$ analytical error bars. The left and right panels are for {\bf Data1} and {\bf Data2} respectively. The upper and lowers curves are  before and after point source subtraction respectively.
 The vertical dotted lines in both the panels show $\ell_{max}$ beyond  which $(\ell > \ell_{max})$
the residual $C_{\ell}$ is dominated by the unsubtracted point sources.} 
\label{fig:fig1}
\end{center}
\end{figure*}

We have used the TGE to estimate $C_{\ell}$ from the measured 
visibilities $\V_i$ with $\u_i$ referring to the corresponding baseline.
As mentioned earlier, the TGE suppresses the contribution from the
residual point sources in the outer region of the telescope's FoV and
also internally subtracts out the noise bias to give an unbiased
estimate of $C_{\ell}$ (details in \citealt{samir14,samir16a}, Paper I).
The tapering is introduced by multiplying the sky with a Gaussian window function
${\cal W}(\theta)=e^{-\theta^2/\theta^2_w}$. The value of
  $\theta_w$ should be chosen in such a way that it cuts off the sky
  response well before the first null of the primary beam without
  removing too much of the signal  from the central region. Here we have
  used $\theta_w=95^{'}$ which is slightly smaller than $114^{'}$, the
  half width at half maxima (HWHM) of the GMRT primary beam  at $150 \, {\rm MHz}$. This is implemented by dividing
the $uv$ plane into a rectangular grid and evaluating the convolved
visibilities $\V_{cg}$ at every grid point $g$ 
\begin{equation}
\V_{cg} = \sum_{i}\tilde{w}(\u_g-\u_i) \, \V_i
\label{eq:a2}
\end{equation}
where $\tilde{w}(\u)$ is the Fourier transform of the taper window
function ${\cal W}(\theta)$ and $\u_g$ refers to the baseline of different grid
points. The entire data  containing visibility measurements in 
different frequency channels that spans a $16 \, {\rm MHz}$ bandwidth 
was collapsed to a single grid after scaling each baseline to the 
appropriate frequency.

The self correlation of the gridded and convolved 
visibilities (equation (10) and (13) of Paper I) can be written as,
\begin{equation}
\begin{split}
\langle \mid \V_{cg} \mid^2 \rangle = \left( \frac{\partial
  B}{\partial T} \right)^2 \int d^2 U \, \mid \tilde{K}\left(\u_g -
\u\right)\,\mid^2 \, C_{2 \pi U_g} \\
+ \sum_i \mid \tilde{w}(\u_g-\u_i)
\mid^2 \langle \mid \N_i \mid^2 \rangle \, ,
\label{eq:a3}
\end{split}
\end{equation}

where, $\left( \frac{\partial B}{\partial T}\right)$ is the conversion factor from brightness temperature to specific intensity,  $\N_i$ is the noise contribution to the individual visibility $\V_i$ and $\tilde{K}\left(\u_g -\u\right)$ is an effective  ``gridding kernel'' which incorporates the effects of (a) telescope's primary beam pattern (b) the tapering window function and (c) the baseline
sampling in the $uv$ plane. 

We have approximated the convolution in equation (\ref{eq:a3}) as,

\begin{equation}
\begin{split}
\langle \mid \V_{cg} \mid^2 \rangle =\Bigg[\left( \frac{\partial B}{\partial T} \right)^2 \int d^2 U \,
\mid \tilde{K}(\u_g-\u) \mid^2 \Bigg] C_{2 \pi U_g} \\
+ \sum_i
\mid \tilde{w}(\u_g-\u_i) \mid^2 \langle \mid \N_i \mid^2 \rangle \, ,
\label{eq:a4}
\end{split}
\end{equation}
under the  assumption that the $C_{\ell}~(\ell=2\pi \mid\u\mid)$ is nearly
constant across the width of $\tilde{K}\left(\u_g - \u\right)$.

We define the Tapered Gridded Estimator (TGE) as
\begin{equation}
{\hat E}_g= M_g^{-1} \, \left( \mid \V_{cg} \mid^2 -\sum_i \mid
\tilde{w}(\u_g-\u_i) \mid^2 \mid \V_i \mid^2 \right) \,.
\label{eq:a6}
\end{equation}

where $M_g$ is the normalizing factor which we have calculated by
using simulated visibilities corresponding to an unit angular power
spectrum (details in Paper I). We have $\langle {\hat E}_g \rangle = C_{\ell_g}$
{\it i.e.} the TGE ${\hat E}_g$ provides an unbiased estimate of the angular power
spectrum $C_{\ell}$ at the angular multipole $\ell_g=2 \pi U_g$ corresponding
to the baseline $\u_g$. We have used the TGE to estimate
$C_{\ell}$ and its variance in bins of equal logarithmic interval in
$\ell$ (equations (19) and (25) in Paper I).

\section{Results and Conclusions}
\label{results}

\begin{figure}
\begin{center}
\psfrag{cl}[b][t][1.5][0]{$C_{\ell} ~~{\rm mK}^2$}
\psfrag{l}[c][c][1.5][0]{$\ell$}
\psfrag{mod3}[r][r][1][0]{$\beta=3.0$}
\psfrag{sim3}[r][r][1][0]{}
\psfrag{mod1.5}[r][r][1][0]{$\beta=1.5$}
\psfrag{sim1.5}[r][r][1][0]{}
\psfrag{data}[r][r][1][0]{Residual}
\psfrag{simu}[r][r][1][0]{Simulation}
\psfrag{model}[r][r][1][0]{$C^M_{\ell}$}
\includegraphics[width=75mm,angle=0]{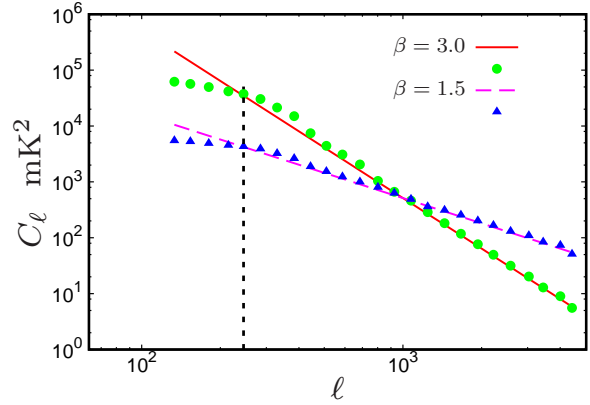}
\caption{Comparison between the estimated $C_{\ell}$ and the model $C^M_{\ell}$ using simulation for {\bf Data1}. The data points and the lines are for the estimated $C_{\ell}$ and the input model $C^M_{\ell}$ (with $\beta=1.5$ and $3$) respectively. We see that the convolution is important 
 in the range $\ell < \ell_{min} = 240$  shown by the vertical dashed  line and we have 
excluded this region from our subsequent analysis. The estimated $C_{\ell}$ matches closely with 
$C^M_{\ell}$ in the range $\ell \ge \ell_{min}$ which we have used for our analysis.} 
\label{fig:fig2}
\end{center}
\end{figure}

\begin{figure*}
\begin{center}
\psfrag{cl}[b][t][1.5][0]{$C_{\ell} ~~{\rm mK}^2$}
\psfrag{l}[c][c][1.5][0]{$\ell$}
\psfrag{mod3}[r][r][1][0]{$\beta=3.0$}
\psfrag{sim3}[r][r][1][0]{}
\psfrag{mod1.5}[r][r][1][0]{$\beta=1.5$}
\psfrag{sim1.5}[r][r][1][0]{}
\psfrag{data}[r][r][1][0]{Residual}
\psfrag{simu}[r][r][1][0]{Simulation}
\psfrag{model}[r][r][1][0]{$C^M_{\ell}$}
\psfrag{50mjy}[r][r][1][0]{$S_c=50{\rm mJy}$}
\includegraphics[width=80mm,angle=0]{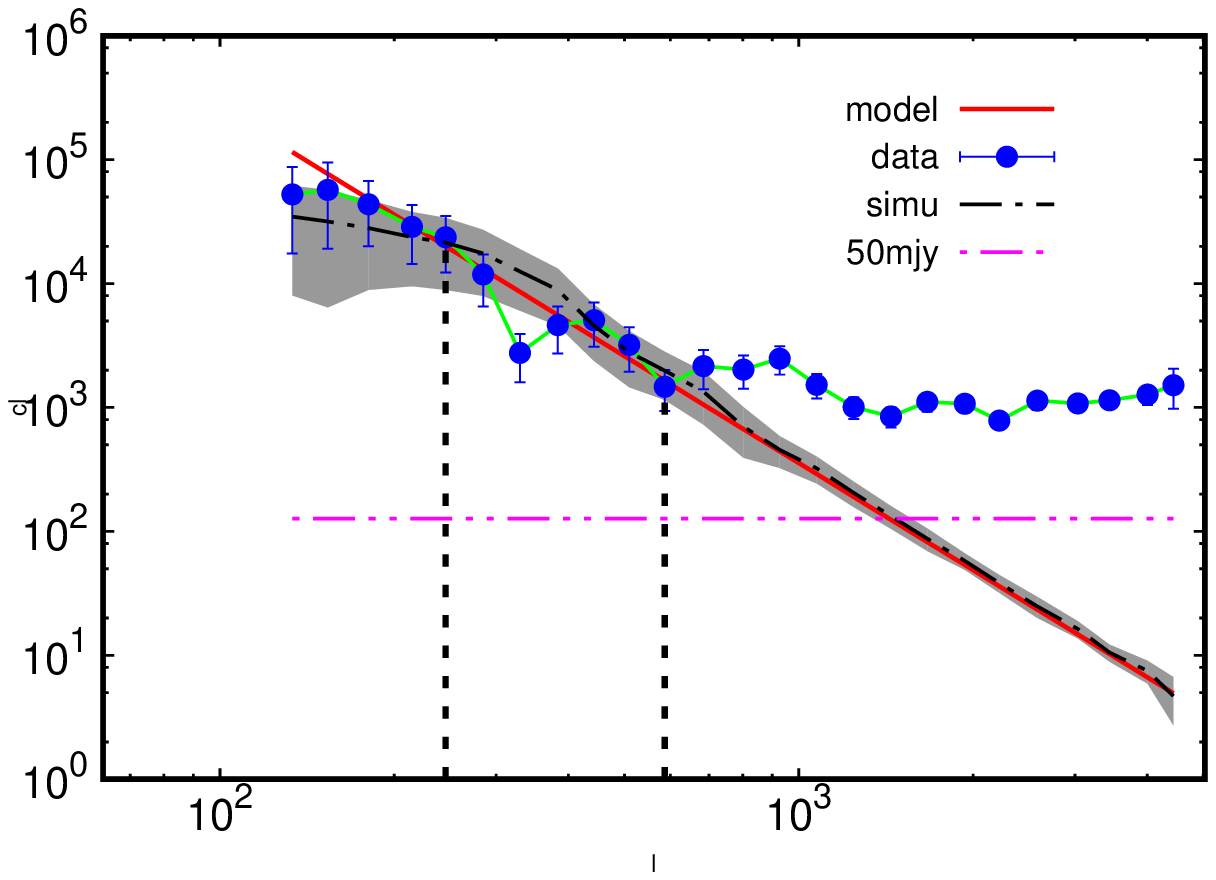}
\includegraphics[width=80mm,angle=0]{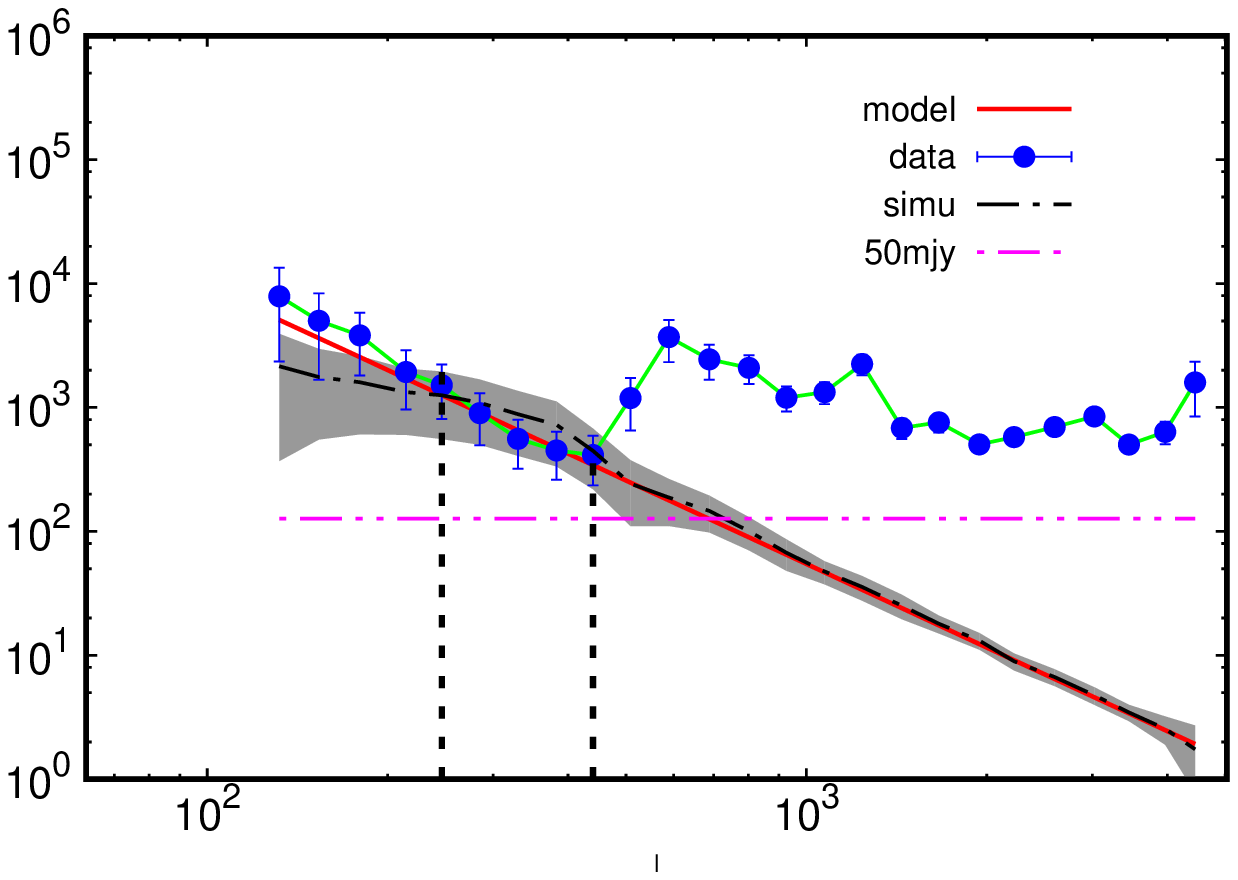}
\caption{Estimated angular power spectra $(C_{\ell})$ using residual data. The left and right panels refer to {\bf Data1} and {\bf Data2} respectively. 
The solid circles with $1-\sigma$ error bars show $C_{\ell}$  estimated  from the residual data,
the vertical dashed lines show the $\ell_{min}-\ell_{max}$ range  used for fitting a power law model 
and the solid lines show the best fit  model.  The dash-dot lines  with the $1-\sigma$ 
shaded region show the mean and standard deviation of $C_{\ell}$ estimated from $128$ realizations 
of simulations with the best fit power law as input model.  The  dot-dot-dash  horizontal 
lines show $C_{\ell}$ predicted from the residual point sources below a threshold flux density 
$S_c=50{\rm mJy}$.  Note that, for {\bf Data2} the estimated values are
only upper limits on the DGSE power spectrum (see Section \ref{results}.)}
\label{fig:fig3}
\end{center}
\end{figure*}

The upper curves of the left and right panels of Figure \ref
{fig:fig1} show the estimated $C_{\ell}$ before point source
subtraction for {\bf Data1} and {\bf Data2} respectively. We find that
for both the data sets the measured $C_{\ell}$ is in the range
$10^4-10^5~{\rm mK^2}$ across the entire $\ell$ range. 
Model predictions \citep{ali08} indicate that the point source contribution
is expected to be considerably larger than the Galactic synchrotron emission
across much of the $\ell$ range considered here, however the two may be
comparable at the smaller $\ell$ values of our interest. Further, the convolution 
in equation (\ref{eq:a3}) is expected to be important at small $\ell$,
and it is necessary to also account for this. 
The lower curves of both the panels of Figure \ref {fig:fig1}
show the estimated $C_{\ell}$ after point source subtraction. We see that
removing the point sources causes a very substantial drop in the 
$C_{\ell}$  measured at large $\ell$. This clearly demonstrates that the 
$C_{\ell}$  at these angular scales was dominated by the point sources prior
to their subtraction. We further believe that after point source subtraction
the $C_{\ell}$  measured at large $\ell$ continues to be dominated by the 
residual point sources   which are below the threshold flux.  The residual flux 
from imperfect subtraction of the bright sources possibly also  makes a
significant contribution in the measured $C_{\ell}$ at large $\ell$. This interpretation is mainly guided by the model
 predictions (Figure 6 of \citealt{ali08}), and is 
also  indicated  by the nearly flat $C_{\ell}$ which is consistent with the
Poisson fluctuations of a random point  source distribution.
 In contrast to
this,  $C_{\ell}$ shows a steep power-law $\ell$ dependence at small
$\ell$ ($\le  \ell_{max}$) with  $\ell_{max}=580$ and $440$ 
for {\bf Data1} and {\bf Data2} respectively. This steep power law is the characteristic of the
diffuse Galactic emission and   we believe that the measured 
$C_{\ell}$ is  possibly dominated by the DGSE at the 
large angular scales corresponding to $\ell \le  \ell_{max}$.
As mentioned earlier,  the convolution  in equation (\ref{eq:a3}) is expected to be 
important at large angular scales  and it is necessary to  account for this in 
order  to correctly interpret the results at small $\ell$.

\begin{table*}
\begin{center}
\begin{tabular}{c c c c c c c c}
\hline
&Galactic Co-ordinate ($l,b$)&$\ell_{min}$&$\ell_{max}$ &A~$({\rm mK^2)}$&$\beta$& $N$ &$\chi^2/(N-2)$\\
\hline
{\bf Data1} & $(9^{\circ},+10^{\circ})$ & $240$ & $580$  & $356\pm109$&  $2.8\pm0.3$ &$6^{a}$ & $0.33$ \\
\hline
{\bf Data2} & $(15^{\circ},-11^{\circ})$ & $240$ & $440$  & $54\pm26$ &  $2.2\pm0.4$ & $5$ & $0.15$\\
\hline
\citealt{bernardi09} & $(137^{\circ},+8^{\circ})$ & $100$ & $900$ & $253\pm40$ & $2.2\pm0.3$ &$-$&$-$\\
\hline
\citealt{ghosh12} & $(151.8^{\circ},+13.89^{\circ})$ & $253$ & $800$ & $513\pm41$ & $2.34\pm0.28$ &$-$&$-$\\
\hline
\citealt{iacobelli13} & $(137^{\circ},+7^{\circ})$ & $100$ & $1300$ & $-$ & $1.84\pm0.19$ &$-$&$-$\\
\hline
 & $(-,\ge +10^{\circ})$ & $-$ & $-$ & ${175}^b$ & $2.88$ &$-$&$-$\\
& $(-,\le-10^{\circ})$ & $-$ & $-$ & ${212}^b$ & $2.74$ &$-$&$-$\\
& $(-,\ge +20^{\circ})$ & $-$ & $-$ & ${85}^b$ & $2.88$ &$-$&$-$\\
\citealt{laporta08} & $(-,\le -20^{\circ})$ & $-$ & $-$ & ${50}^b$ & $2.83$ &$-$&$-$\\
& $(-,\ge +10^{\circ})$ & $-$ & $-$ & ${691}^c$ & $2.80$ &$-$&$-$\\
& $(-,\le-10^{\circ})$ & $-$ & $-$ & ${620}^c$ & $2.70$ &$-$&$-$\\
& $(-,\ge +20^{\circ})$ & $-$ & $-$ & ${275}^c$ & $2.83$ &$-$&$-$\\
& $(-,\le-20^{\circ})$ & $-$ & $-$ & ${107}^c$ & $2.87$ &$-$&$-$\\
\hline
\end{tabular}
\caption{This shows the values of the parameters which are used to fit the data. In comparison, the parameters from other observations are also shown in this table.  For {\bf Data2}, the best fit values are derived with the assumption that the residual contribution is negligible below ${\ell_{max}}$.
\newline
 ${}^a$ Excluding one outlier point; ${}^b$ Extrapolated from {\rm 1420 MHz} to {\rm 147.5 MHz}; ${}^c$ Extrapolated from {\rm 408 MHz} to {\rm 147.5 MHz.}}
\label{tab:1}
\end{center}
\end{table*}

We have carried out simulations in order to assess the effect of the 
convolution on the estimated  $C_{\ell}$. GMRT visibility data was simulated 
assuming that the sky brightness temperature fluctuations are a realization of 
a Gaussian random field with input model angular power spectrum $C^M_{\ell}$ of 
the form given by  eq. (\ref{eq:eq1}). The simulations incorporate the GMRT primary beam pattern and the $uv$ tracks corresponding to the actual observation  under consideration. 
The reader is referred to \citet{samir14}  for more details of the simulations. 
Figure \ref {fig:fig2} shows the $C_{\ell}$ estimated from
the {\bf Data1} simulations  for  $\beta =3$ and $1.5$ which roughly
encompasses the entire range of  the 
power law index we expect for the Galactic synchrotron emission. 
We find that the effect of the
convolution is important in the range  $\ell<\ell_{min}=240$, and we have
excluded this $\ell$ range from our analysis.  
We are, however,  able to recover the input model angular
power spectrum quite accurately in the region $\ell\ge\ell_{min}$ which we
have used for our subsequent analysis.  We have also carried out the same
analysis for  {\bf  Data2} (not shown here) where we find that $\ell_{min}$
has a value that is almost the same as for  {\bf  Data1}.

We have used the $\ell$ range $\ell_{min} \le \ell \le \ell_{max}$ to fit a power law of the form 
given in eq. (\ref{eq:eq1}) to the $C_{\ell}$ measured  after point source subtraction. 
The data points with $1-\sigma$ error bars and the best fit power law are shown in 
 Figure \ref {fig:fig3}. Note that we have identified one of the {\bf Data1} points as 
an outlier and excluded it from the fit. The best fit parameters $(A,\beta)$, 
$N$ the number  of data points used for the fit and $\chi^2/(N-2)$ the 
chi-square  per degree of freedom (reduced $\chi^2$) are listed in Table ~\ref{tab:1}.  
The rather low values of the reduced $\chi^2$ indicate that the errors in the measured 
$C_{\ell}$ have possibly been somewhat overestimated.
In order to
validate our methodology we have simulated the visibility data for an
input model power spectrum with the best fit values of the parameters
$(A,\beta)$ and used this to estimate $C_{\ell}$. The mean $C_{\ell}$
and $1-\sigma$ errors (shaded region) estimated from $128$ realization
of the simulation are shown in Figure \ref {fig:fig3}. For the
relevant $\ell$ range we find that the simulated $C_{\ell}$ is in very
good agreement with the measured values thereby validating the entire
fitting procedure.  The horizontal lines in both the panels of Figure
\ref {fig:fig3} show the $C_{\ell}$ predicted from the Poisson
fluctuations of residual point sources below a threshold flux density
of $S_c=50~{\rm mJy}$.  The $C_{\ell}$ prediction here is based on the
$150 \, {\rm MHz}$ source counts of \citet{ghosh12}.  We find that for
$\ell>\ell_{max}$ the measured $C_{\ell}$ values are well in excess of
this prediction indicating that (1.) there are significant residual
imaging artifacts around the bright source ($S > S_c$) which were
subtracted , and/or (2.) the actual source distribution is in excess
of the predictions of the source counts. Note that the actual $S_c$
values ($20.5$ and $15.5 \, {\rm mJy}$ for {\bf Data1} and {\bf Data2}
respectively) are well below $50 \, {\rm mJy}$, and the corresponding
$C_{\ell}$ predictions will lie below the horizontal lines shown in
Figure \ref {fig:fig3}.

 For both the fields $C_{\ell}$ (Figure \ref {fig:fig3}) is
  nearly flat  at large $\ell$ $(> 500)$  and  it  is well modeled 
  by a power law  at smaller $\ell$ ($240 \le \ell \lesssim 500$). For 
{\bf Data1}   the  power law rises above the flat $C_{\ell}$, and the
power law is likely  dominated by the DGSE. However, for {\bf Data2}
the  power law  falls below the flat $C_{\ell}$,
and it is likely that in addition to the DGSE there is a  significant
residual point sources contribution. 
For {\bf Data2} we interpret the best fit power law as an
upper limit for the DGSE.

The best fit parameters  $(A,\beta)=(356.23\pm109.5,2.8\pm0.3)$ and
$(54.6\pm26,2.2\pm0.4)$ for {\bf Data1} and {\bf Data2} respectively 
are compared with  measurements from other $150 \, {\rm MHz}$ observations such as \citet{bernardi09,ghosh12,iacobelli13}  in Table~\ref{tab:1}. Further, we have also used an earlier work (\citealt{laporta08}) at higher frequencies $(\rm 408$ and $\rm 1420~MHz)$ to estimate and compare the amplitude of the angular power spectrum of the DGSE  expected at our observing frequency.
Using the best-fit parameters (tabulated at $\ell$ = 100) at 408 and 1420 MHz, we extrapolate the amplitude of the $C_{\ell}$  at our observing frequency at $\ell = 1000$  for $|b| \ge 10^{\circ}$ and  $|b| \ge 20^{\circ}$. In this extrapolation we use a mean frequency spectral index of $\alpha = 2.5$ (\citealt{costa}) $(C_{\ell}\propto\nu^{2\alpha})$. The extrapolated amplitude values are shown in Table ~\ref{tab:1}.  In Table ~\ref{tab:1}, we note that the angular power spectra of  the DGSE  in  the northern hemisphere are comparatively larger than  that of the southern hemisphere. The best fit parameter $A$ for  {\bf Data1}({\bf Data2})  agrees mostly with the extrapolated values obtained from $b \ge +10^{\circ}$ ($b \le -10^{\circ}$) and  $b \ge +20^{\circ}$ ( $b \le -20^{\circ}$) within a factor of about 2 (4). The best fit parameter $\beta$ for  {\bf Data1} and {\bf Data2} is within the range of 1.5-3.0 found by all the previous measurements at ${\rm 150~MHz}$ and  higher frequencies.

 The entire analysis here is based on the assumption that the 
 DGSE is a Gaussian random field. This is possibly justified for the
 small patch of the sky under observation given that the  diffuse
 emission is generated by a random processes like  MHD turbulence. The estimated $C_{\ell}$ remains
unaffected even if this assumption breaks down, only the error
estimates will be changed. We  note that the  parameters  $(A,\beta)$ are varying significantly from field to field across the different direction in the sky. We plan to extend this analysis for the whole sky and study the variation of the amplitude $(A)$ and power law index $(\beta)$ of $C_{\ell}$ using the full TGGS survey in future.

\section{Acknowledgements}
We thank an anonymous referee for helpful comments. S. Choudhuri would like to acknowledge the University Grant Commission, India for providing financial support. AG would like acknowledge Postdoctoral Fellowship from the South African Square Kilometre Array Project for financial support. We thank the staff of the GMRT that made these observations possible. GMRT is run by the National Centre for Radio Astrophysics of the Tata Institute of Fundamental Research. We thank the staff of the GMRT that made these observations possible. GMRT is run by the National Centre for Radio Astrophysics of the Tata Institute of Fundamental Research.

\end{document}